\documentclass[preprint2]{aastex}

\usepackage{natbib}
\usepackage{graphics}
\usepackage{amsmath,mathtools}
\usepackage{multirow}

\defcitealias{AliHaimoud:2009p322}{AHD09}
\defcitealias{Draine:1998p126}{DL98}
\defcitealias{Hoang:2010p316}{HDL10}
\defcitealias{Hoang:2011p269}{HLD11}
\defcitealias{Silsbee:2011p317}{SAH11}
\defcitealias{Weingartner:2001p526}{WD01}

\shorttitle{Analytical Spectrum of Spinning Dust}
\shortauthors{Stevenson, M.}

\begin{document}

\title{Derivation of an Analytical Approximation of the Spectrum of Spinning Dust Emission}


\author{Matthew A. Stevenson}
\affil{California Institute of Technology, Mail Code 249-17, 
    Pasadena, CA 91125}
\email{mas@astro.caltech.edu}

\begin{abstract}
An analytical function for the spectrum of spinning dust emission is presented.  It is derived through the application of careful approximations, with each step tested against numerical calculations.  This approach ensures accuracy while providing an intuitive picture of the physics.  The final result may be useful for fitting of anomalous microwave emission observations, as is demonstrated by a comparison with the Planck observations of the Perseus Molecular Cloud.  It is hoped that this will lead to a broader consideration of the spinning dust model when interpreting microwave continuum observations, and that it will provide a standard framework for interpreting and comparing the variety of anomalous microwave emission observations.
\end{abstract}



\section{Introduction}

The existence of an anomalous component of diffuse microwave emission is well established, though it has yet to mature as an astrophysical probe.  This emission was first detected as a cosmological foreground by \citet{Kogut:1996p672} and \citet{deOliveiraCosta:1997p673}, first being discovered to be anomalous by \citet{Leitch:1997p119} in observations near the North Celestial Pole.  It was quickly demonstrated by \citet[\citetalias{Draine:1998p126} hereafter]{Draine:1998p126} to be consistent with electric dipole radiation from very small dust grains, a process now commonly referred to as spinning dust emission.  This explanation has gained wide favor, though perhaps prematurely, as it remains to be proven that this is the cause of the North Celestial Pole emission.  Free-free emission from very hot gas and magnetic dipole emission from dust grains \citep{Draine:1999p124, Draine:2013p674} may plausibly explain the anomalous emission in this region and in some others.

Anomalous microwave emission has now been observed by many authors in a variety of Galactic and extragalactic environments \citep{Finkbeiner:2002p796,Finkbeiner:2004p797,Murphy:2010p211,Lu:2012p675,Murphy:2012p726}.  The emission is characterized by a broad peak around $20-40\,\mathrm{GHz}$, spatial correlation with dust on degree scales, peak brightness roughly four orders of magnitude less than that of thermal dust emission, and little polarization \citep{RubinoMartin:2012p676}.  Puzzlingly, a strong correlation with infrared tracers of small grains at arcminute scales has not been observed \citep{Tibbs:2011p677, Tibbs:2012p678}.

The original model of \citetalias{Draine:1998p126} was derived under the key assumptions of a Maxwellian distribution of grain angular velocity, grain rotation about the axis of maximum moment of inertia, simple grain geometries (spherical, disk-like, and rod-like), and electric dipole moments of the grains based on random walks over chemical bonds.  This model found wide success in fitting anomalous microwave emission measurements.  \citet{Finkbeiner:2004p125} and \citet{Gold:2009p504} are notable examples of this.

\citet{Lazarian:2000p802} explored the theory of polarized spinning dust radiation, finding that the radiation could not be polarized by more than $10\,\%$, and even then only below $10\,\mathrm{GHz}$.  Since then, observational studies have consistently found upper limits of anomalous microwave emission polarization at the percent level \citep{Dickinson:2007p801,Mason:2009p799,Macellari:2011p800,RubinoMartin:2012p676}.  These upper limits have been interpreted to support the spinning dust model.  Most recently, \citet{Hoang:2013p798} used the $2175\,\mathrm{\AA}$ polarization feature, as observed for two stars, to argue that the spinning dust polarization should peak at $3\,\%$ at $5\,\mathrm{GHz}$, and decrease rapidly above $20\,\mathrm{GHz}$.

Since \citetalias{Draine:1998p126}, spinning dust theory has advanced in both precision and scope.  \citet{Rafikov:2006p679} applied the theory to protoplanetary disks, while \citet{Ysard:2010p527} showed that a quantum treatment gave the same results as the classical approach of \citetalias{Draine:1998p126}.  \citet[\citetalias{AliHaimoud:2009p322} hereafter]{AliHaimoud:2009p322} advanced the theory by allowing for non-Maxwellian distributions of grain rotation via the Fokker-Planck equation, through refined treatments of the excitation and damping processes, and by producing the SpDust IDL package, which allowed users to calculate custom spectra given astrophysical parameters.  \citet[\citetalias{Hoang:2010p316} hereafter]{Hoang:2010p316}  considered the dramatic effects of irregular rotation about non-principal axes and used the Langevin equation instead of the Fokker-Planck equation so to capture the transient effects due to collisions with individual ions.  \citet[\citetalias{Silsbee:2011p317} hereafter]{Silsbee:2011p317} updated SpDust to include irregular rotation and improved calculations of the rotational damping and excitation.  \citet[\citetalias{Hoang:2011p269} hereafter]{Hoang:2011p269} extended the theory to irregularly shaped grains and further explored the distribution of rotational energies arising from vibrational-rotational energy coupling.  These refinements and extensions have been accompanied by increases in complexity: the latest models depend on upwards of 30 parameters.

SpDust has had a large impact on this field.  It allows for quick calculation of models using nine physical parameters and has shown great utility in fitting observations \citep{Planck:2011p338}.   Its use has been limited, however, in cosmological foreground separation.  Nine parameters is more than the shape of the spectrum justifies, and the code runs too slowly to allow rapid exploration of multi-dimensional parameter space.

Foreground separation efforts have instead resorted to simple, analytical models with three parameters or less.  No single function has found wide use.  \citet{Bonaldi:2007p513}, following \citet{Tegmark:1998p681} and \citet{deOliveiraCosta:2004p682}, suggested a parabola in $\log S-\log\nu$ space, \citet{Tegmark:2000p680} put forth a modified graybody, and \citet{Gold:2009p504} simply shifted the numerical models of \citetalias{Draine:1998p126}.  Although expedient for fitting, these approaches do not easily lead to astrophysical interpretation.  An analytical function that is easily relatable to the physics would offer an advantage:  it would be well suited to fitting and to interpretation.  It is the aim of this paper to provide such a function by analytically deriving the spinning dust spectrum through use of careful approximations.

The approach in this study is to follow the derivation of \citetalias{Silsbee:2011p317}, but using analytical approximations where numerical calculations would otherwise be required.  The approximations are tested against the results of SpDust to demonstrate where they succeed in capturing the numerical model.  As this approach uses the Fokker-Planck equation instead of the Langevin equation, it is not possible to reproduce the transient spin-up effects of \citetalias{Hoang:2010p316}.  These effects were shown to be contained in the high-frequency fall-off the spectrum, and their exclusion does represent an inaccuracy in this new approach.  The triaxial grains and range of vibrational-rotational energy coupling considered in \citetalias{Hoang:2011p269} are not directly addressed in this paper, although the present treatment of irregular rotation is extensible to such effects.  These omissions should be considered when applying this model to data.  Polarization of the spinning dust radiation is not considered.

This paper is structured as follows.  An overview of the derivation steps is provided in Section~\ref{overview.sec}.  Section~\ref{dust.sec} describes the assumed dust properties.  The rotational distribution function, and its dependence on environment, is discussed in Section~\ref{distribution.sec}.  Section~\ref{emissivity.sec} presents the emissivity itself, including a treatment of irregular rotation.  Finally, Section~\ref{discussion.sec} provides a discussion of the derived function, its use, and various caveats.

\section{Overview}\label{overview.sec}

The total emissivity of an ensemble of rotating grains, $j_\nu/n_H$, is the integral of the emissivity of grains of a given size, $j_\nu^a$, weighted by the grain size distribution, $1/n_H\,dn_\mathrm{gr}/da$.  This is written as
\begin{equation}\label{grain_integral_general.eq}
\frac{j_\nu}{n_H} = \frac{1}{n_H}\int_{a_\mathrm{min}}^{a_\mathrm{max}}da\frac{dn_\mathrm{gr}}{da}j_\nu^a.
\end{equation}
The grain emissivity is calculated by integrating emitted power over the angular momentum and electric dipole moment distributions,
\begin{equation}\label{single_grain_general.eq}
j_\nu^a = 2\pi\int_0^\infty4\pi J^2dJ\int_0^\infty d\mu f_a(J,\mu)P(\mu)\frac{P_\mathrm{ed,\omega}(J,\mu)}{4\pi},
\end{equation}
in which $f_a(J,\mu)$ is the angular momentum distribution function for grains of size $a$ and electric dipole moment $\mu$, $P(\mu)$ is the electric dipole moment distribution,  and $P_\mathrm{ed,\omega}(J,\mu)$ is the power emitted at frequency $\omega$.  This last function accounts for the complex, torque-free motion of aspherical grains (called ``wobbling'' in \citetalias{Hoang:2010p316} and ``tumbling'' in \citetalias{Silsbee:2011p317}).

The integrals benefit from two changes of variable.  The first is to calculate rotation using the ratio of angular momentum to maximum moment of inertia
\begin{equation}\label{Omega_def.eq}
\Omega \equiv \frac{J}{I_M}
\end{equation}
rather than the angular momentum itself.  $\Omega$ is henceforth referred to as the rotation rate, though it is understood that this label is only truly accurate in the non-tumbling case.  The second is to separate the electric dipole moment from the grain size using a new variable $b$ (as discussed in Section~\ref{dipolemoment.sec}).  Equation~\ref{single_grain_general.eq} then becomes
\begin{equation}\label{j_nu_a_integral.eq}
j_\nu^a = 2\pi \int_0^\infty \Omega^2d\Omega\int_0^\infty db f_a(\Omega,b)P(b)P_{\mathrm{ed},\omega}(\Omega,b).
\end{equation}

The strategy adopted in this paper is to make a number of judicious simplifications aimed at approximating Equation~\ref{j_nu_a_integral.eq} as a log-normal function.  Equation~\ref{grain_integral_general.eq} is then evaluated analytically to give the desired result.

A large number of symbols are used in this paper.  For the convenience of the reader, the most important of these are compiled in Table~\ref{variables.tab}.

\begin{deluxetable}{llc}
\tablecolumns{3}
\tablewidth{0pt}
\tabletypesize{\footnotesize}
\tablecaption{Important Variables Used in This Paper.
\label{variables.tab}}
\tablehead{ 
Variable    &  Description &  Equation} 
\startdata 
$a$ & Grain size & (\ref{n_atoms.eq}) \\
$N_\mathrm{at}$ & Number of atoms & (\ref{n_atoms.eq}, \ref{grain_size_dist.eq}) \\
$\frac{1}{n_H}\frac{dn_\mathrm{gr}}{da}$ & Distribution of $a$ & (\ref{grain_size_dist.eq}) \\
$B_1$ & Normalization of $a$ distribution & (\ref{grain_size_dist.eq}) \\
$a_m$ & Minimum grain size & (\ref{grain_size_dist.eq}) \\
$a_0$ & Peak grain size & (\ref{grain_size_dist.eq}) \\
$\sigma$ & Width of $a$ distribution & (\ref{grain_size_dist.eq}) \\[6pt]
$b$ & Normalized electric dipole moment & (\ref{beta_def.eq})\\
$\epsilon_\mathrm{ip}$ & In-plane $b$ fraction & (\ref{epsilon_def.eq})\\
$P\left(b\right)$ & Distribution of $b$ & (\ref{p_beta_general.eq}) \\
$\beta$ & Width of $b$ distribution & (\ref{p_beta_general.eq}) \\
$N_b$ & Dimension of $b$ distribution & (\ref{p_beta_general.eq}) \\[6pt]
$\Omega$ & Grain rotation rate & (\ref{Omega_def.eq})\\
$\omega=2\pi\nu$ & Radiation frequency & (\ref{omega_def.eq})\\
$q_r$ & Ratio of $\omega$ to $\Omega$ & (\ref{omega_def.eq})\\[6pt]
$f_a\left(\Omega,b\right)$ & $\Omega$ distribution function & (\ref{fokker_planck.eq})\\
$A_\Omega$ & Exp. coefficient of $\Omega$ distribution & (\ref{plaw_approx.eq}) \\
$\alpha_a$ & Power law on $a$ & (\ref{plaw_approx.eq}) \\
$\alpha_b$ & Power law on $b$ & (\ref{plaw_approx.eq}) \\
$\alpha_\nu$ & Power law on $\Omega$ & (\ref{plaw_approx.eq}) \\
$\Omega_{p,a}$ & Peak $\Omega$ for size $a$ & (\ref{Omega_p_a_def.eq}) \\[6pt]
$\mathcal{I}_a\left(\Omega\right)$ & $\Omega$ distribution, integrated over $\beta$ & (\ref{I_a_def.eq}, \ref{I_a_approx.eq}) \\
$\mathcal{I}_0$ & Normalization of $\mathcal{I}_a$ & (\ref{I_0_def.eq}) \\
$\sigma_\Omega$ & Width of $\mathcal{I}_a$ & (\ref{sigma_Omega_def.eq}) \\[6pt]
$P_{\mathrm{ed},\omega}\left(\Omega,b\right)$ & Emission from single grain & (\ref{P_ed_omega.eq}) \\
$R\left(\omega,\Omega\right)$ & Dimensionless emission spectrum & (\ref{P_ed_omega.eq}, \ref{R_approx.eq}) \\
$R_0$ & Normalization of $R$ & (\ref{R_approx.eq}) \\
$\sigma_r$ & Width of $R$ & (\ref{R_approx.eq}) \\[6pt]
$j_\nu^a$ & Emissivity for grains of size $a$ & (\ref{j_nu_a_integral.eq}, \ref{j_nu_a_redef.eq}, \ref{j_nu_a_approx.eq}) \\
$\sigma_\nu$ & Width of $j_\nu^a$ & (\ref{sigma_nu_def.eq}) \\[6pt]
$j_\nu/n_H$ & Total emissivity per H & (\ref{grain_integral_general.eq}, \ref{j_nu_nh.eq}) \\
$\alpha_s$ & Power law of $j_\nu/n_H$ & (\ref{alpha_s.eq}) \\
$\sigma_s$ & Log-normal width of $j_\nu/n_H$ & (\ref{sigma_s.eq}) \\
$\nu_0$ & Characteristic frequency of $j_\nu/n_H$ & (\ref{nu_0.eq}) \\
$\eta_\nu$ & Error function slope for $\nu$ & (\ref{eta_nu.eq}) \\
$\eta_a$ & Error function slope for $a_m$ & (\ref{eta_a.eq}) \\ \noalign{\smallskip}
\enddata
\end{deluxetable}

\section{Dust Grains}\label{dust.sec}

Spinning dust emission is sensitive to fundamental properties of the grains.  The grain sizes and permanent electric dipole moments are the most important; geometry and charge are of lesser concern.  A simple, thermal calculation shows that rotation at tens of GHz requires sub-nm grains.  Such a population is consistent with the polycyclic aromatic hydrocarbon population described in \citet[\citetalias{Weingartner:2001p526} hereafter]{Weingartner:2001p526} and \citet{Draine:2007p524}, although debate persists regarding the relative importance of aliphatic and aromatic structures in these grains \citep{Kwok:2011p683}.  This population is needed to explain the observed infrared emission and its properties can be constrained by observations of ultraviolet extinction.

\subsection{Size}

If $a$ is the spherical-equivalent radius, then the number of atoms per grain is roughly
\begin{equation}\label{n_atoms.eq}
N_\mathrm{at} \approx 600 \left(\frac{a}{1\,\mathrm{nm}}\right)^3,
\end{equation}
consistent with the prescription of  \cite{Li:2001p528} if there is one hydrogen atom for every three carbons.  The grains of interest thus contain fewer than $600$ atoms.  The smallest may be plausibly described as large molecules.

A log-normal size distribution is conventionally assumed for these grains \citep[\citetalias{Weingartner:2001p526};][]{Compiegne:2011p684}.  As noted by \citetalias{Weingartner:2001p526}, this form is not motivated by physics, but by mathematical convenience.  This distribution is accompanied by a second log-normal distribution peaking at $3\,\mathrm{nm}$ and a power-law extending beyond $0.1\,\mathrm{\mu m}$, though these additional components are insignificant below $1\,\mathrm{nm}$.  Inspired by photolytic considerations \citep{Guhathakurta:1989p536}, the distribution is assumed to truncate sharply at a smallest grain size.  I therefore approximate the size distribution as
\begin{equation}\label{grain_size_dist.eq}
\frac{1}{n_H}
\frac{dn_\mathrm{gr}}{da} = 
\begin{dcases}
0 & a < a_\mathrm{m} \\
\frac{B_1}{a}\exp\left\{-\frac{1}{2}\left[\frac{\log\left(a/a_0\right)}{\sigma}\right]^2\right\} & a \geq a_\mathrm{m}.
\end{dcases}
\end{equation}
Following \citetalias{Weingartner:2001p526} and \citetalias{AliHaimoud:2009p322}, the values $B_1=1.2\times10^{-6}$, $\sigma=0.4$, and $a_0=3.5\,\mathrm{\AA}$ are used when calculating model parameters, though in practice these can be varied if the data require.  In particular, $B_1$ represents the abundance of the small grains and there is no reason to expect it to be fixed by nature.  Breaking from previous approaches, I do not assume that $a_0$ and $a_\mathrm{m}$ are equal.

The log-normal form of Equation~\ref{grain_size_dist.eq} heavily influenced the mathematics of this paper.  Different size distributions would require different approximations to be made in Sections~\ref{tumbling.sec} and \ref{grain_emissivity.sec}, resulting in a qualitatively different analytical forms for $j_\nu/n_H$.

\subsection{Shape}

It is unlikely that these grains have simple shapes.  \citetalias{Draine:1998p126} assumed rod-like and disk-like geometries for the smallest grains, inspired by aliphatic and aromatic molecules.  Spherical shapes were assumed for larger grains.  Sharp transitions between these occur at sizes $a_1$ and $a_2$, with the grains smaller than $a_1$ being rod-like, grains larger than $a_1$ but smaller than $a_2$ being disk-like, and grains larger than $a_2$ being spherical.  Later models followed \citetalias{Draine:1998p126} in setting $a_2=6\,\mathrm{\AA}$, but set $a_1=\,0$.  These precedents are followed here, though it is shown in Section~\ref{total_emissivity.sec} that $a_2$ has little effect on the final result.

Grain shape influences the rotational distribution functions by way of grain cross sections, charge distributions, and electric dipole moment geometry.  These effects are felt in the numerically calculated values of the parameters of Equation~\ref{plaw_approx.eq} and when considering the irregular rotation of grains.  In Section~\ref{tumbling.sec}, the effects of geometry and irregular rotation are parameterized as part of the full derivation.  This parameterization is applicable to the range of plausible geometries, although only disk-like and spherical grains are explicitly considered.

\subsection{Temperature}

The internal temperatures of the grains are not constant.  The grains are transiently heated by UV photon absorption and cool near to ground state before the next UV photon is absorbed.  The result is a grain temperature distribution \citep{Guhathakurta:1989p536}.  This is important for the rotational distribution functions, as it will affect IR photon emission rate and the atom desorption rates and evaporation temperatures.

The temperatures of the grains will be coupled to their rotational energy.  \citetalias{Hoang:2010p316} and \citetalias{Hoang:2011p269} showed that the strength of this coupling, or the rate of internal relaxation, has a significant effect on the grain tumbling.  In the case of strong coupling, there will be a minimum vibrational temperature at which coupling can occur (due to the sparsity of the vibrational mode spectrum at low temperatures).  If this temperature is much greater than the rotational energy of the grain, then there will be a uniform distribution of $\sin\theta$, where $\theta$ is the rotation angle: the angle between the grain's angular momentum and axis of maximum moment of inertia.  This is the case considered by \citetalias{Silsbee:2011p317}.  Conversely, if the decoupling temperature is much less than rotational energy, then the rotation angle will be zero.  When coupling is weak, the rotation angle is governed by a Maxwellian distribution.  Section~\ref{tumbling.sec} explicitly calculates the grain tumbling in the  case of strong coupling with a high decoupling temperature, though the suggested parameterization can also be applied to the other cases.

\subsection{Charge}

Collisional and photoelectric charging of grains has implications for electric dipole moments and interaction cross-sections.  \citetalias{Draine:1998p126} considered this and presented the charge distribution functions for a variety of grain sizes and environments.  Sub-nm grains had typical charges between $-1$ and $3$.  Such small charges are unlikely to dominate the electric dipole moments.  They are, however, important when calculating grain rotation rates (see Section~\ref{distribution.sec}).

\subsection{Dipole Moment}\label{dipolemoment.sec}

The intrinsic, electric dipole moments of the grains are poorly constrained observationally, and attempts to derive them theoretically are subject to uncertainty in the specific chemical compositions of the grains.  \citetalias{Draine:1998p126} instead assumed a typical moment per molecular bond $b$ and used a random walk over all bonds to get the total dipole moment $\mu$.  \citetalias{AliHaimoud:2009p322} extended this by having $\mu$ normally distributed.  The variance is then
\begin{equation}\label{mu_def.eq}
\left<\mu^2\right> = N_\mathrm{at}\beta^2.
\end{equation}
Note that this distribution is a function of grain size.

With the aim of separating the integrals over grain size and dipole moment cleanly, I have taken a different approach.  Defining the normalized dipole moment, $b$, via
\begin{equation}\label{beta_def.eq}
b^2 \equiv \frac{\mu^2}{N_\mathrm{at}}
\end{equation}
allows use of the normal distribution
\begin{equation}\label{p_beta_general.eq}
P(b) =\frac{2\left(\frac{N_b}{2}\right)^\frac{N_b}{2}}{\Gamma\left(\frac{N_b}{2}\right)} \frac{1}{b}\left(\frac{b}{\beta}\right)^{N_b}
\exp\left[-\frac{N_b}{2}\left(\frac{b}{\beta}\right)^2\right].
\end{equation}
$N_b$ is the dimensionality of the distribution and is 1, 2, or 3 for linearly, cylindrically, and spherically distributed dipole moments.  $\Gamma(x)$ is the Gamma function.  When calculating numerical parameters, $\beta$ is taken as $0.4\,\mathrm{D}$.

\citetalias{Silsbee:2011p317} considered the case of disk-like grains with three-dimensional electric dipole moment distributions, which could be due to disk warping from pentacyclene structures.  They parameterize this possibility via the in-plane fraction of the dipole moment:
\begin{equation}\label{ip_def.eq}
\epsilon_\mathrm{ip} = \frac{\left<\mu_\mathrm{ip}^2\right>}{\left<\mu^2\right>}.
\end{equation}
The out-of-plane fraction is similarly defined, and
\begin{equation}\label{epsilon_def.eq}
\epsilon_\mathrm{ip} + \epsilon_\mathrm{op}= 1.
\end{equation}

\section{Distribution Function}\label{distribution.sec}

The physics of dust grain rotation is nontrivial.  Desired is a rotational distribution function, which will be a function of grain size and astrophysical environment.  Smaller grains tend to rotate faster due to smaller moments of inertia, while for a given grain size, the preferred rotation rate is the result of a variety of excitation and damping mechanisms.

It is useful to think of rotation rate using an intuitive picture, in which the various excitation and damping mechanisms are competing to thermalize grain rotation to their respective temperatures.  Torques from the emission of infrared photons (which follow thermal spikes due to UV photon absorption) push the grain rotation towards the average IR radiation temperatures, which depend upon the grain heat capacities and emission spectra and can reach $\sim\,10^3\,\mathrm{K}$.  Desorption of atoms (adsorbed via gas collisions) pushes to the evaporation temperature, of order $10^2\,\mathrm{K}$.  Plasma interactions cause the rotation to tend to the gas temperature, which varies widely with interstellar phase.  At the same time, drag from the electric dipole emission itself can limit grain rotation, causing the distribution function to fall off non-thermally at high rotation rates.

Detailed treatments of these effects need to be done numerically, and no attempt to reproduce or improve upon these efforts are made in this paper.  See \citetalias{Draine:1998p126}, \citetalias{AliHaimoud:2009p322}, \citetalias{Hoang:2010p316}, and \citetalias{Silsbee:2011p317} for careful discussions and calculations.  Rather, I will show how a simple parameterization of the distribution function can encompass the important effects.  

\citetalias{AliHaimoud:2009p322} and \citetalias{Silsbee:2011p317} used the Fokker-Planck equation to calculate the distribution function.  This differential equation allows one to account for damping and excitation of a stationary system from small impulses.  Adapted from \citetalias{Silsbee:2011p317},
\begin{equation}\label{fokker_planck.eq}
\frac{df_a\left(\Omega,b\right)}{d\Omega} + \frac{I_M\Omega^2}{kT}\frac{F}{G} \frac{f_a\left(\Omega,b\right)}{\Omega} = 0,
\end{equation}
where
\begin{equation}\label{F_sum.eq}
F=\sum_jF_j
\end{equation}
and
\begin{equation}\label{G_sum.eq}
G=\sum_jG_j
\end{equation}
are the sums of the dimensionless damping and excitation coefficients.  Note that, while \citetalias{Silsbee:2011p317} treated the electric dipole damping as a separate term, it is included here as one of the $F_j$.  In the current work, the above is simplified further by assuming that the actions of dipole moment, grain size, and rotation frequency are separable and are described by power laws (with influences of ISM environment, grain charge, and grain temperature being folded into the parameterization):
\begin{equation}\label{plaw_approx.eq}
\frac{I_M\Omega^2}{kT}\frac{F}{G} \equiv
\alpha_\nu
A_\Omega
\left(\frac{b}{\beta}\right)^{\alpha_b}
\left(\frac{a}{a_0}\right)^{\alpha_a}
\left(\frac{\Omega}{\Omega_{p,a_0}}\right)^{\alpha_\nu}.
\end{equation}
In quantifying these assumptions, this equation serves as the definition of the power-law indices and the peak rotation frequency $\Omega_{p,a_0}$ for grains of size $a_0$.  $\Omega_{p,a_0}$ is guaranteed to be the peak frequency of the rotational distribution function by the definition of $A_\Omega$,
\begin{equation}\label{A_Omega_def.eq}
\begin{array}{l}
A_\Omega^2 \equiv
\left(\frac{N_b}{2}\right)^{\alpha_b} \\
\frac{
\left[
\Gamma(8/\alpha_\nu)
\Gamma(N_b/2+1-5\alpha_b/2\alpha_\nu)
\right]^{8\alpha_\nu}
}{
\left[
\Gamma(7/\alpha_\nu)
\Gamma(N_b/2+1-2\alpha_b/\alpha_\nu)
\right]^{5\alpha_\nu}
\left[
\Gamma(9/\alpha_\nu)
\Gamma(N_b/2+1-3\alpha_b/\alpha_\nu)
\right]^{3\alpha_\nu}
}.
\end{array}
\end{equation}
Although quite useful, it should be clear that reducing the Fokker-Planck equation to this form may introduce degeneracy amongst the astrophysical parameters and ultimately limit the physics one can infer when fitting this model.

The power law indices and rotational peak can be acquired directly from Equation~\ref{plaw_approx.eq} given numerically calculated tables of the $F_j$ and $G_j$.  SpDust was used to do this for the idealized environments of \citetalias{Draine:1998p126}: cold neutral medium (CNM), dark cloud (DC), molecular cloud (MC), photodissociation region (PDR), reflection nebula (RN), warm ionized medium (WIM), and warm neutral medium (WNM).  The results of these calculations are presented in Table~\ref{alpha_allsize.tab}.  For each environment, the parameters are calculated for disk-like (1) and spherical (2) grains at sizes of $4.5\,\mathrm{\AA}$ and at $6.3\,\mathrm{\AA}$.  The peak rotation frequencies are extrapolated to $a_0$ and $a_2$.  

\begin{deluxetable}{lc@{\hspace{30pt}}rrrrr@{\hspace{20pt}}rrrrr}
\tablecolumns{12}
\tablewidth{0pt}
\tablecaption{Rotational distribution function parameters.
\label{alpha_allsize.tab}}
\tablehead{ 
\colhead{} & \colhead{} & \multicolumn{10}{c}{Parameter} \\ \cline{3-12} 
\colhead{} & \colhead{} & \multicolumn{4}{c}{Calculated at $4.5\,\mathrm{\AA}$} & \colhead{} & \multicolumn{4}{c}{Calculated at $6.3\,\mathrm{\AA}$} & \colhead{} \\ \cline{3-6} \cline{8-11} 
 \multicolumn{2}{l}{Environment} & \colhead{$\alpha_a$} & \colhead{$\alpha_b$} & \colhead{$\alpha_\nu$} & \colhead{$\Omega_{p,a}$} & \colhead{$\Omega_{p,a_0}$} & \colhead{$\alpha_a$} & \colhead{$\alpha_b$} & \colhead{$\alpha_\nu$} & \colhead{$\Omega_{p,a}$} & \colhead{$\Omega_{p,a_2}$}} 
\startdata
 \multirow{2}{*}{CNM}  & (1)  &   5.01 &   1.63 &   3.86 &   98.6 &  135.3 &   5.49 &   1.32 &   3.36 &   60.9 &   66.1 \\ 
   & (2)  &   5.09 &   1.54 &   3.93 &  153.3 &  210.3 &   5.43 &   1.29 &   3.58 &   94.8 &  102.3 \\[4pt] 
 \multirow{2}{*}{DC}  & (1)  &   5.69 &   0.12 &   2.10 &   70.9 &  137.1 &   6.10 &   0.06 &   2.01 &   26.8 &   31.3 \\ 
   & (2)  &   4.56 &   0.14 &   2.13 &  115.4 &  194.4 &   4.60 &   0.11 &   2.04 &   54.4 &   61.0 \\[4pt] 
 \multirow{2}{*}{MC}  & (1)  &   5.05 &   1.59 &   3.32 &  126.5 &  183.3 &   5.93 &   1.47 &   2.58 &   67.2 &   75.4 \\ 
   & (2)  &   3.53 &   1.57 &   3.36 &  200.9 &  259.7 &   5.63 &   1.60 &   2.83 &  124.8 &  137.9 \\[4pt] 
 \multirow{2}{*}{PDR}  & (1)  &   6.66 &   0.13 &   2.13 &  367.6 &  787.4 &   6.67 &   0.01 &   2.01 &  120.6 &  142.5 \\ 
   & (2)  &   5.95 &   0.25 &   2.25 &  653.2 & 1243.2 &   5.79 &   0.03 &   2.03 &  253.8 &  292.9 \\[4pt] 
 \multirow{2}{*}{RN}  & (1)  &   6.19 &   0.31 &   2.33 &  200.5 &  383.3 &   6.23 &   0.04 &   2.03 &   74.0 &   86.4 \\ 
   & (2)  &   5.20 &   0.52 &   2.54 &  355.5 &  584.9 &   5.31 &   0.10 &   2.10 &  161.3 &  183.1 \\[4pt] 
 \multirow{2}{*}{WIM}  & (1)  &   4.77 &   1.69 &   3.97 &   93.4 &  125.2 &   4.39 &   1.60 &   3.79 &   62.6 &   66.4 \\ 
   & (2)  &   4.92 &   1.57 &   4.01 &  146.6 &  197.8 &   4.99 &   1.44 &   3.88 &   95.1 &  101.5 \\[4pt] 
 \multirow{2}{*}{WNM}  & (1)  &   4.28 &   1.73 &   3.95 &   81.6 &  106.3 &   4.21 &   1.58 &   3.71 &   56.2 &   59.4 \\ 
   & (2)  &   4.62 &   1.64 &   3.99 &  127.9 &  169.7 &   4.80 &   1.50 &   3.82 &   84.4 &   89.9 \\ 
\enddata
\tablecomments{(1): Disk-like grains. (2): Spherical grains.} 
\end{deluxetable}

\begin{figure}[htb]
\begin{center}
\plotone{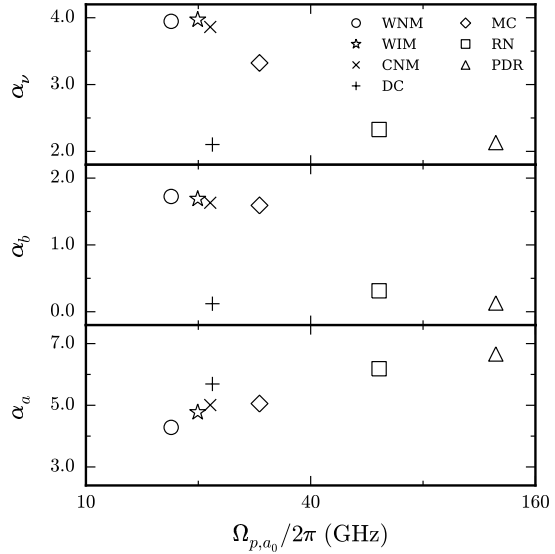}
\caption{Power law parameters from Equation~\ref{plaw_approx.eq} plotted against peak rotation frequency for the ideal interstellar environments.  Parameters are calculated numerically from SpDust.  Data are listed in Table~\ref{alpha_allsize.tab}.}
\label{fa_params.fig}
\end{center}
\end{figure}

These parameters are plotted in Figure~\ref{fa_params.fig}.  If the astrophysical parameters were not already degenerate prior to imposing Equation~\ref{plaw_approx.eq}, then the correlations between these parameters ought to be weak.  As can be seen, however, the parameters are highly correlated.   The dichotomy seen in $\alpha_b$ and $\alpha_\nu$ reflects the cases of whether or not electric dipole damping is dominant, with $\alpha_b\approx0$ and $\alpha_\nu\approx2$ in the latter case.  The correlation between $\alpha_a$ and $\Omega_{p,a_0}$ is more subtle, as the various $F_j$ and $G_j$ depend on $a$ to different degrees.  The result of these correlations is the implication that the astrophysical parameters are themselves highly degenerate, and that inferring environmental physics from this emission would be challenging even without the power law approximation.

It is useful to define, $\Omega_{p,a}$, the peak rotation frequency for grains of size $a$,
\begin{equation}\label{Omega_p_a_def.eq}
\left(\frac{a}{a_0}\right)^{\alpha_a}
\left(\frac{\Omega_{p,a}}{\Omega_{p,a_0}}\right)^{\alpha_\nu} \equiv 1.
\end{equation}
Equation~\ref{plaw_approx.eq} becomes
\begin{equation}\label{plaw_approx_a.eq}
\frac{I_M\Omega^2}{kT}\frac{F}{G} =
\alpha_\nu
A_\Omega
\left(\frac{b}{\beta}\right)^{\alpha_b}
\left(\frac{\Omega}{\Omega_{p,a}}\right)^{\alpha_\nu}
\end{equation}
and the Fokker-Planck equation is integrated to give
\begin{equation}\label{fa_equation.eq}
\begin{array}{l}
f_a(\Omega,b) =
\frac{\alpha_\nu A_\Omega^{3/\alpha_\nu}}{4 \pi\Gamma(3/\alpha_\nu)\Omega_{p,a}^3}
\left(\frac{b}{\beta}\right)^{3\alpha_b/\alpha_\nu} \\
\exp
\left[-
A_\Omega
\left(\frac{b}{\beta}\right)^{\alpha_b}
\left(\frac{\Omega}{\Omega_{p,a}}\right)^{\alpha_\nu}
\right].
\end{array}
\end{equation}

\begin{figure}[htbp]
\begin{center}
\plotone{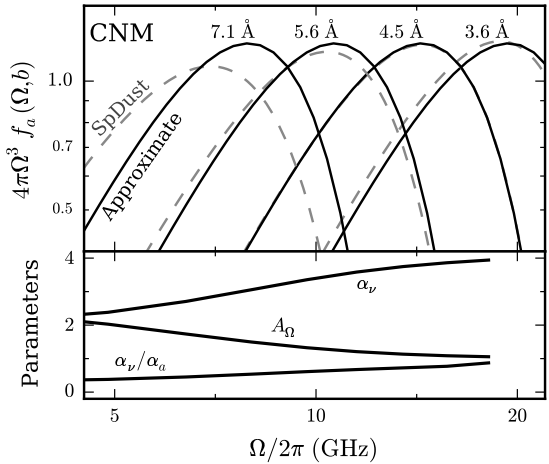}
\caption{Rotational distribution functions for grains of various sizes in the CNM environment with $b=\beta$ are shown in the upper panel.  Solid, black curves show the analytical, power law approximation.  Gray, dashed curves show the numerically calculated functions from SpDust.  Lower panel shows the variation of power law parameters with grain size.}
\label{fa_vs_omega.fig}
\end{center}
\end{figure}

The validity of this approach is demonstrated in Figure~\ref{fa_vs_omega.fig}.  Distribution functions for a variety of grain sizes are plotted for the CNM environment.  Curves calculated from SpDust are compared to the results of the power law approximation (extrapolated from $4.5\,\mathrm{\AA}$ grains).  Agreement is satisfactory for grains smaller than $6\,\mathrm{\AA}$, but then deteriorates rapidly.  This disagreement is due to variations in $\alpha_a$ and $\alpha_\nu$ (indicating a failure of the power law approximation), also shown in this figure.  This figure demonstrates that the power law approximation is reasonable over small ranges in grain size, but becomes a significant source of error when used over wider ranges of sizes and frequencies.  In particular, one should be careful when using parameters derived at $a_0=3.5\,\mathrm{\AA}$ when $a_m>6\,\mathrm{\AA}$.

\section{Emissivity}\label{emissivity.sec}

It is now possible to seek a solution to the integrals in Equations~\ref{grain_integral_general.eq} and \ref{j_nu_a_integral.eq}.  Further approximations will be needed in order to achieve an analytical result, and a log-normal form is suggested by the grain size distribution.  These integrals are approached with this goal in mind.  I first consider the effects of grain tumbling, then find a log-normal approximation for $j_\nu^a$, and finally complete the derivation of $j_\nu/n_H$.

In the following, tumbling is only considered in the case of axisymmetric grains with strong vibrational-rotational coupling and high decoupling temperatures, resulting in a uniform distribution in the Sine of the rotation angle.  Although these techniques may also be applied to triaxial grains and the broader variety of vibrational-rotational coupling, these cases are not treated here.

\subsection{Grain Tumbling}\label{tumbling.sec}

The effects of irregular grain rotation manifest themselves in the power emitted by a single grain.  In general, this can written as
\begin{equation}\label{P_ed_omega.eq}
P_{\mathrm{ed},\omega}\left(\Omega,b\right) =
\mathcal{P}_{p,a}
\frac{b^2}{\beta^2}\frac{\Omega^4}{\Omega_{p,a}^4}R\left(\omega,\Omega\right)
\end{equation}
where
\begin{equation}\label{P_p_a.eq}
\mathcal{P}_{p,a} \equiv \frac{2}{3}\frac{\beta^2N_\mathrm{at}\omega_{p,a}^4}{c^3}.
\end{equation}
This form naturally allows for the various permutations of grain geometry and rotation dynamics, with the emission spectrum itself being contained in the $R$ function.  The emission frequency is related to rotation frequency by
\begin{equation}\label{omega_def.eq}
\omega \equiv q_r\Omega
\end{equation}
and the emission spectrum is described by $R(\omega,\Omega)$.

\begin{figure}[htbp]
\begin{center}
\plotone{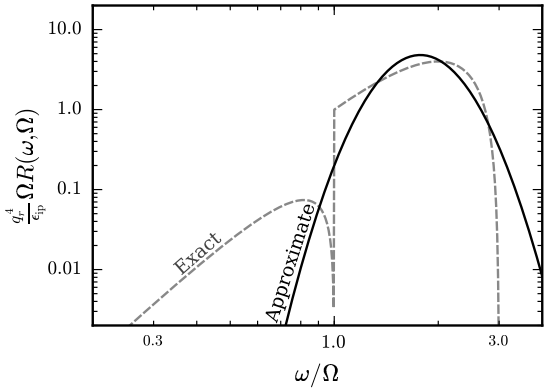}
\caption{Exact vs. approximate forms of the tumbling spectrum for in-plane emission from disk-like grains.}
\label{R_vs_omega.fig}
\end{center}
\end{figure}

In the non-tumbling case, $q_r=1$.  The spectrum is a delta function:
\begin{equation}\label{R_disk_notumbling.eq}
R\left(\omega,\Omega\right) = \epsilon_\mathrm{ip}\delta\left(\omega - \Omega\right)
\end{equation}
or
\begin{equation}\label{R_sphere.eq}
R\left(\omega,\Omega\right) = \frac{2}{3}\delta\left(\omega - \Omega\right)
\end{equation}
for disk-like and spherical grains, respectively.  In the case of tumbling, there is emission due to the in-plane and out-of-plane electric dipole moments.  From \citetalias{Silsbee:2011p317}, the out-of-plane emission has $q_r=2$ and
\begin{equation}\label{R_disk_outofplane.eq}
R\left(\omega,\Omega\right) = \frac{2\epsilon_\mathrm{op}}{3}\delta\left(\omega - 2\Omega\right)
\end{equation}
while in-plane has
\begin{equation}
R\left(\omega,\Omega\right) = 
\begin{dcases} \frac{\epsilon_\mathrm{ip}}{4q_r^4}\frac{\omega^4}{\Omega^5}\left(3-\frac{\omega}{\Omega}\right)^2 & \Omega<\omega<3\Omega \\
\frac{\epsilon_\mathrm{ip}}{2q_r^4}\frac{\omega^4}{\Omega^5}\left(1-\frac{\omega^2}{\Omega^2}\right) &  \omega<\Omega
\end{dcases}.
\end{equation}
The latter case does not lend itself to analytical progress, so it is approximated it with a log-normal function having the same first and second moments.  This approximation is shown in Figure~\ref{R_vs_omega.fig}.  The fit is clearly not perfect, yet it deviates by less than $10\,\%$ of the peak across most of the range.  
The approximated function is
\begin{equation}\label{R_approx.eq}
R(\omega,\Omega) \approx \frac{R_0}{\sqrt{2\pi}\sigma_r \omega} \exp\left\{-\frac{1}{2}\left[\frac{\log{(\omega/q_r\Omega)}-\sigma_r^2}{\sigma_r}\right]^2\right\}
\end{equation}
with integral
\begin{equation}\label{R0_def.eq}
R_0 = \frac{5\epsilon_\mathrm{ip}}{q_r^4}
\end{equation}
width
\begin{equation}\label{sigma_r.eq}
\sigma_r^2 \approx 0.0518
\end{equation}
and peak
\begin{equation}
q_r \approx 1.775.
\end{equation}

Triaxial grains, lower vibrational-rotational coupling temperatures, and weak vibrational-rotational coupling, as described in 
\citetalias{Hoang:2010p316} and \citetalias{Hoang:2011p269}, are not explicitly considered here.  The above approximation can be applied to these cases, resulting in different values for $q_r$, $\sigma_r$, and $R_0$.

It is convenient to continue the derivation using the log-normal form of $R(\omega,\Omega)$.  The results can be applied to non-tumbling cases by setting $\sigma_r=0$ and $R_0$ equal to the coefficients in Equations~\ref{R_disk_notumbling.eq}, \ref{R_sphere.eq}, and \ref{R_disk_outofplane.eq}.

\subsection{Grain Emissivity}\label{grain_emissivity.sec}

The integrals of Equation~\ref{j_nu_a_integral.eq} can be rewritten as
\begin{equation}\label{j_nu_a_redef.eq}
j_\nu^a = \frac{1}{2}\mathcal{P}_{p,a} \int_0^\infty d\Omega R\left(\omega,\Omega\right)
\mathcal{I}_a\left(\Omega\right)
\end{equation}
in which
\begin{equation}\label{I_a_def.eq}
\mathcal{I}_a(\Omega) = 4\pi\int_0^\infty db\frac{b^2}{\beta^2}P(b)
\frac{\Omega^6 f_a\left(\Omega,b\right)}{\Omega_{p,a}^4}.
\end{equation}
In this, the functions $\mathcal{I}_ad\Omega$ represents the dimensionless rotation spectrum and $R$ the dimensionless emission spectrum for a given rotation rate.  Equation~\ref{j_nu_a_redef.eq} is a simple convolution of these.  $R$ has already been approximated as a log-normal function, so if a similar approximation to $\mathcal{I}_a$ can be found, then  $j_\nu^a$ will have the desired form.

\begin{figure}[htbp]
\begin{center}
\plotone{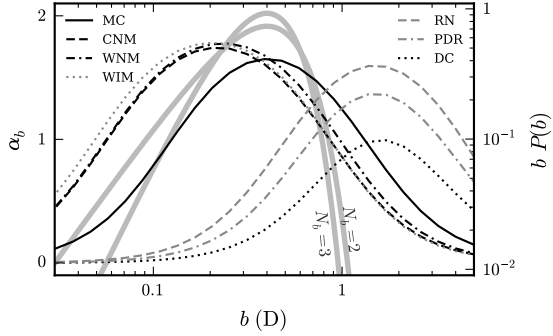}
\caption{$\alpha_b$ vs $b$ for $3.5\,$\AA\ grains at peak emission frequency.  Curves are calculated via SpDust.  Light grey curves show $b\,P(b)$ for $N_b$ of 2 and 3.}
\label{alpha_beta_vs_beta.fig}
\end{center}
\end{figure}

The main concern is whether the assumed power law $\alpha_b$ is constant over the $b$ integral.  A nonzero $\alpha_b$ indicates that grains of different dipole moments will rotate at different frequencies, so deviation from the power-law assumption leads to errors in the width and peak frequency of this integral.  Figure~\ref{alpha_beta_vs_beta.fig} shows $\alpha_b$ for the ideal environments and demonstrates that the power-law assumption is reasonable, as the $\alpha_b$ do not change greatly over the peak of $P(b)$.  

A log-normal approximation for $\mathcal{I}_a(\Omega)$ is achieved by calculating the first and second moments of $\Omega$ over Equation~\ref{I_a_def.eq}.  The integral over $b$ becomes analytical once $\Omega$ has been integrated, giving the result
\begin{equation}\label{I_a_approx.eq}
\mathcal{I}_a(\Omega) \approx \frac{\mathcal{I}_0}{\sqrt{2\pi}\sigma_\Omega}\frac{1}{\Omega} \exp\left\{-\frac{1}{2}\left[\frac{\log{(\Omega/\Omega_{p,a})}-\sigma_\Omega^2}{\sigma_\Omega}\right]^2\right\}
\end{equation}
with
\begin{equation}\label{I_0_def.eq}
\mathcal{I}_0 = \left(\frac{N_b}{2}\right)^{2\alpha_b/\alpha_\nu}
\frac{\Gamma(7/\alpha_\nu) \Gamma(N_b/2+1-2\alpha_b/\alpha_\nu)}
{\Gamma(N_b/2+1)\Gamma(3/\alpha_\nu)}A_\Omega^{-4/\alpha_\nu}
\end{equation}
and
\begin{equation}\label{sigma_Omega_def.eq}
\begin{array}{l}
\sigma_\Omega^2 = \\
\log\left[
\frac{\Gamma(7/a_\nu) \Gamma(N_b/2+1- 2\alpha_b/\alpha_\nu) \Gamma(9/\alpha_\nu)\Gamma(N_b2+1-3\alpha_b/\alpha_\nu)}{ \Gamma(8/\alpha_\nu)^2\Gamma(N_b/2+1-5\alpha_b/2\alpha_\nu)^2}
\right].
\end{array}
\end{equation}

\begin{figure}[htbp]
\begin{center}
\plotone{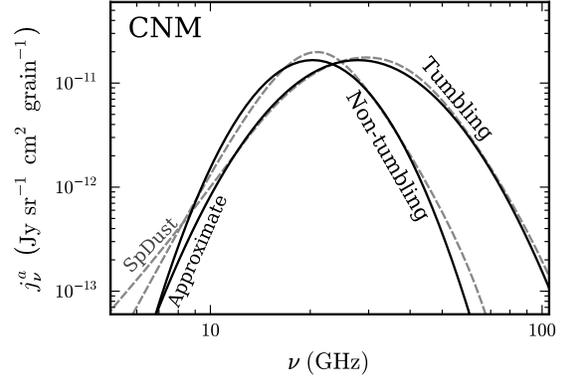}
\caption{Emissivity for an ensemble of grains of size $3.5\,\mathrm{\AA}$ in the CNM environment.  Disk-like grains are assumed in the tumbling and non-tumbling cases.  Black, solid curves are analytically approximated.  Gray, dashed curves are numerically calculated from SpDust.}
\label{j_nu_a_vs_nu.fig}
\end{center}
\end{figure}

The emissivity for grains of size $a$ then follows immediately:
\begin{equation}\label{j_nu_a_approx.eq}
j_\nu^a \approx  \frac{\mathcal{P}_{t,a}}{4\pi}
\frac{1}{ \sqrt{2 \pi}\sigma_\nu\nu}
\exp\left\{-\frac{1}{2}\left[\frac{\log{(\nu/\nu_{p,a})} - \sigma_\nu^2}{\sigma_\nu}\right]^2\right\}
\end{equation}
in which
\begin{equation}\label{sigma_nu_def.eq}
\sigma_\nu^2 = \sigma_r^2 + \sigma_\Omega^2
\end{equation}
and
\begin{equation}\label{P_t_a.eq}
\mathcal{P}_{t,a}  = \frac{2}{3}\frac{\beta^2N_\mathrm{at}\omega_{p,a}^4}{c^3}R_0\mathcal{I}_0.
\end{equation}
This is plotted in Figure~\ref{j_nu_a_vs_nu.fig} for disk-like grains in both the tumbling and non-tumbling cases.  The analytical functions continue to show satisfactory agreement with the numerically derived curves.

\subsection{Total Emissivity}\label{total_emissivity.sec}

Given the above approximations, the integration over grain size follows analytically,
\begin{equation}\label{j_nu_nh.eq}
\begin{array}{l}
\frac{j_\nu}{n_H} =
\left.\frac{j_\nu}{n_H}\right|_{\nu_0}
\left(\frac{\nu}{\nu_0}\right)^{\alpha_s}
\exp\left\{-\frac{1}{2}\left[\frac{\log{(\nu/\nu_0)}}{\sigma_s}\right]^2\right\} \\
\mathrm{erfc}\left[
\eta_\nu\log\frac{\nu}{\nu_0} + 
\eta_a\log\frac{a_m}{a_0}
\right].
\end{array}
\end{equation}
The characteristic frequency
\begin{equation}\label{nu_0.eq}
\nu_0 \equiv \nu_{p,a_0}\exp\left(-\alpha_s\sigma_\nu^2\right)
\end{equation}
is that at which grains of size $a_0$ make their greatest fractional contribution to the total emissivity (assuming a flat grain size distribution).  The power law and log-normal width are
\begin{equation}\label{alpha_s.eq}
\alpha_s \equiv 3 - 3\frac{\alpha_\nu}{\alpha_a}
\end{equation}
and 
\begin{equation}\label{sigma_s.eq}
\sigma_s^2 \equiv \frac{\alpha_a^2}{\alpha_\nu^2}\sigma^2 + \sigma_\nu^2.
\end{equation}
The emissivity at $\nu_0$ is 
\begin{equation}\label{j_nu_nh_0.eq}
\left.\frac{j_\nu}{n_H}\right|_{\nu_0}  =B_1 
A_s
\frac{\beta^2N_\mathrm{at}\omega_{0}^3}{6c^3}
\end{equation}
where
\begin{equation}\label{A_s.eq}
A_s  \equiv
\frac{\sigma}{\sigma_s}R_0\mathcal{I}_0\exp\left[-\frac{\sigma_\nu^2}{2}\left(9\frac{\alpha_\nu^2}{\alpha_a^2}-8\right)\right].
\end{equation}
The complementary error function provides the high-frequency fall-off with parameters
\begin{equation}\label{eta_nu.eq}
\eta_\nu \equiv \frac{1}{\sqrt{2}}\frac{\alpha_a\sigma}{\alpha_\nu\sigma_\nu\sigma_s}
\end{equation}
and
\begin{equation}\label{eta_a.eq}
\eta_a \equiv \frac{1}{\sqrt{2}}\frac{\sigma_s}{\sigma_\nu\sigma}.
\end{equation}

\begin{figure}[htbp]
\begin{center}
\plotone{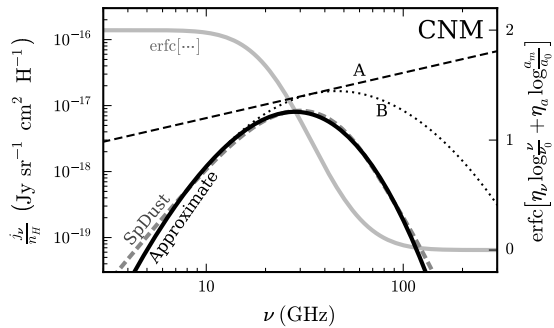}
\caption{Total spinning dust emissivity for the CNM environment.  Solid, black curve is the analytical function of Equation~\ref{j_nu_nh.eq}, while the dashed, gray curve is that from SpDust.  Components of the analytical curve are also shown:  $A$ is the power-law term and $B$ is the power-law multiplied by the log-normal distribution.  The high-frequency fall-off is provided by the complementary error function, shown in gray using the right $y-$axis.}
\label{j_nu_vs_nu_cnm.fig}
\end{center}
\end{figure}

\begin{figure*}[htbp]
\begin{center}
\includegraphics[width=6in]{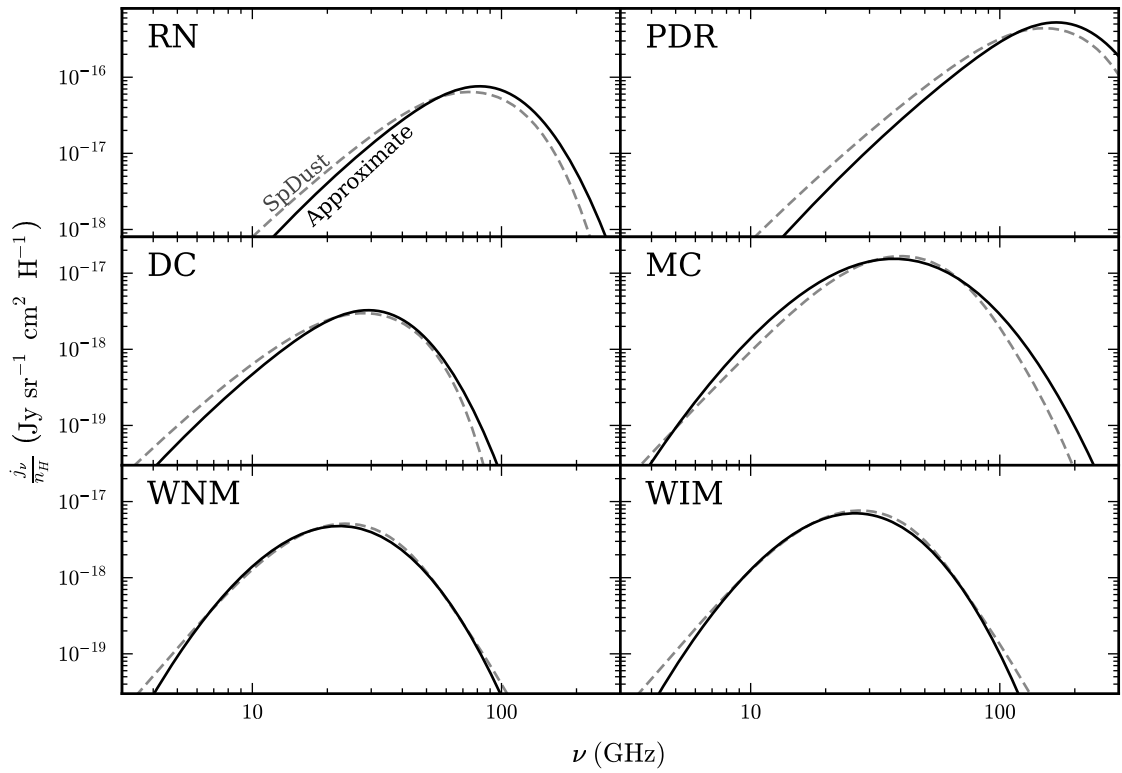}
\caption{Analytical estimates (solid, black) of $j_\nu/n_H$ compared to that of SpDust (gray, dashed) for idealized interstellar environments.}
\label{j_nu_vs_nu_all.fig}
\end{center}
\end{figure*}

Equation~\ref{j_nu_nh.eq} is the chief result of this paper.  It should be regarded as the natural, functional form for the spinning dust emission.  The components of this function are plotted in Figure~\ref{j_nu_vs_nu_cnm.fig}, in which the power-law and log-normal components are shown in turn, as is the complementary error function.  The analytical curve is plotted alongside the numerically calculated emissivity from SpDust.  The agreement is excellent.  This is particularly noteworthy as the parameter $a_2$ has been disregarded: the grain geometry is thus of only secondary importance.  The same is shown for the rest of the ideal environments in Figure~\ref{j_nu_vs_nu_all.fig}.

\begin{figure}[htbp]
\begin{center}
\plotone{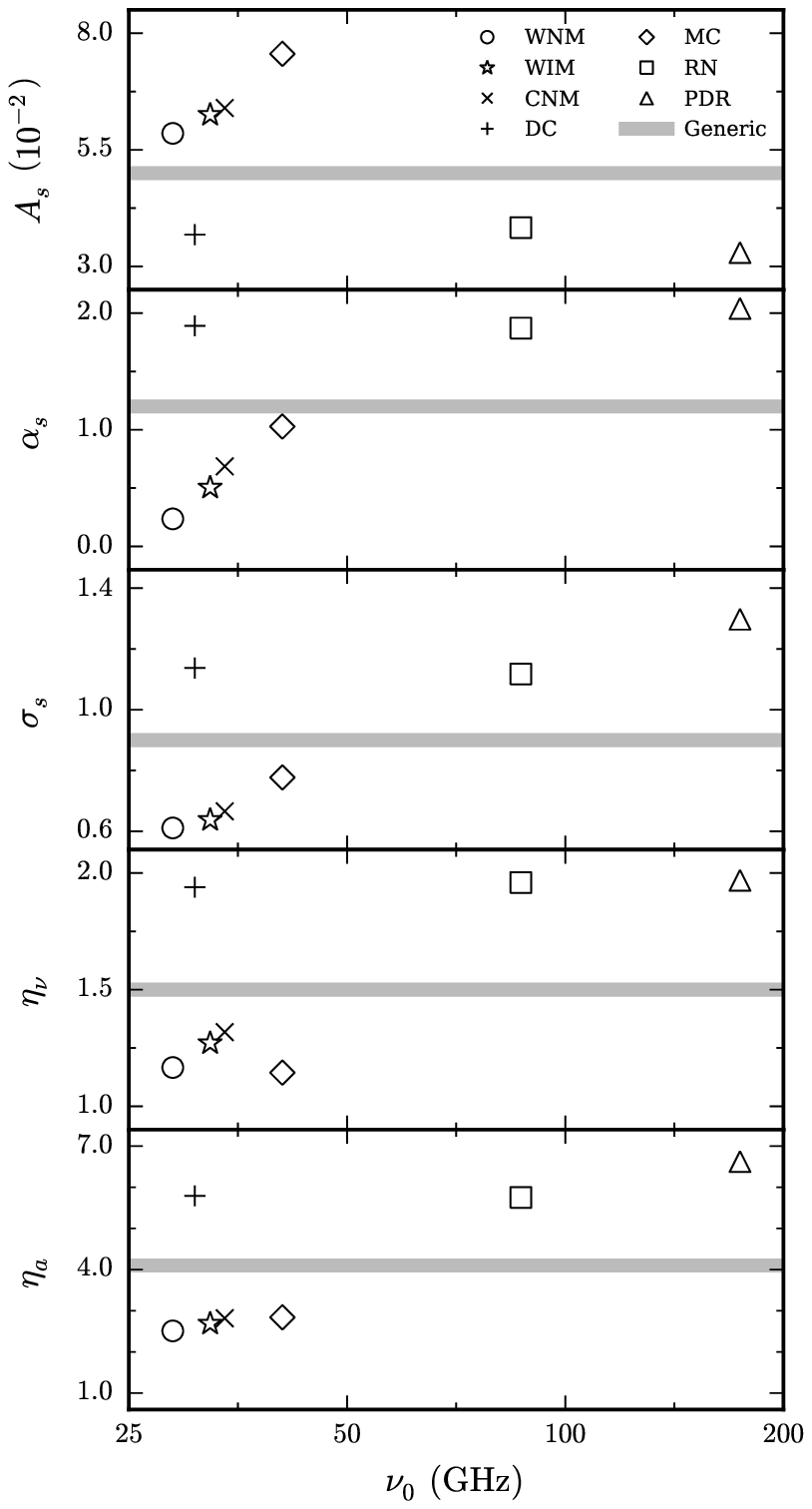}
\caption{Parameters of the analytical $j_\nu/n_H$ function for the idealized interstellar environments. The suggested generic parameters are shown by the gray, horizontal lines.}
\label{j_nu_params.fig}
\end{center}
\end{figure}

The function contains six parameters, which are heavily degenerate both in derivation and effect.  These are the amplitude coefficient $A_s$, the characteristic frequency $\nu_0$, the power-law slope $\alpha_s$, the log-normal width $\sigma_s$, and the exponential slopes $\eta_\nu$ and $\eta_a$.  The parameters have been calculated in Table~\ref{alpha_derived.tab} for the idealized environments for disk-like (1) and spherical (2) grains.  The same is plotted in Figure~\ref{j_nu_params.fig} for case (1).  Also shown are a generic set of parameters recommended for use when, for example, the data are not able to break degeneracies between the parameters.  Indeed, the strong correlations between these parameters are clear.

\begin{deluxetable}{lc@{\hspace{30pt}}crrrrrr}
\tablecolumns{12}
\tablewidth{0pc}
\tablecaption{Derived parameters for the analytical emissivity.
\label{alpha_derived.tab}}
\tablehead{ 
\colhead{} & \colhead{} & \multicolumn{7}{c}{Parameter} \\ \cline{3-9} 
\colhead{} & \colhead{} & \colhead{$\left.\frac{j_\nu}{n_H}\right|_{\nu_0}$} & \colhead{$\nu_0$} & \colhead{$A_s$} & \colhead{$\alpha_s$} & \colhead{$\sigma_s$} & \colhead{$\eta_\nu$} & \colhead{$\eta_a$} \\ 
 \multicolumn{2}{l}{Environment} &  \colhead{$\left(\mathrm{Jy\,sr^{-1}\,cm^{-2}\,H^{-1}}\right)$} &  \colhead{$\left(\mathrm{GHz}\right)$} &  \colhead{$\left(10^{-2}\right)$}& \colhead{} & \colhead{} & \colhead{} & \colhead{} } 
\startdata
 Generic & & \nodata & \nodata  &   5.00 &  1.200 &  0.900 &   1.50 &   4.10 \\[4pt] 
 \multirow{2}{*}{CNM}  & (1)  & $7.48\times10^{-18}$ &   33.9 &   6.40 &  0.686 &  0.666 &   1.32 &   2.82 \\ 
   & (2)  & $7.77\times10^{-18}$ &   31.8 &   8.08 &  0.685 &  0.587 &   2.26 &   3.76 \\[4pt] 
 \multirow{2}{*}{DC}  & (1)  & $3.24\times10^{-18}$ &   30.8 &   3.68 &  1.892 &  1.137 &   1.94 &   5.79 \\ 
   & (2)  & $3.26\times10^{-18}$ &   27.8 &   5.08 &  1.599 &  0.895 &   2.59 &   6.07 \\[4pt] 
 \multirow{2}{*}{MC}  & (1)  & $1.53\times10^{-17}$ &   40.7 &   7.56 &  1.028 &  0.777 &   1.14 &   2.84 \\ 
   & (2)  & $1.56\times10^{-17}$ &   40.7 &   7.71 &  0.148 &  0.533 &   1.71 &   2.88 \\[4pt] 
 \multirow{2}{*}{PDR}  & (1)  & $5.23\times10^{-16}$ &  174.2 &   3.30 &  2.040 &  1.297 &   1.97 &   6.63 \\ 
   & (2)  & $6.98\times10^{-16}$ &  175.0 &   4.34 &  1.864 &  1.087 &   2.68 &   7.48 \\[4pt] 
 \multirow{2}{*}{RN}  & (1)  & $7.53\times10^{-17}$ &   86.8 &   3.83 &  1.871 &  1.117 &   1.96 &   5.75 \\ 
   & (2)  & $9.92\times10^{-17}$ &   84.3 &   5.52 &  1.532 &  0.856 &   2.65 &   5.94 \\[4pt] 
 \multirow{2}{*}{WIM}  & (1)  & $6.36\times10^{-18}$ &   32.4 &   6.25 &  0.505 &  0.638 &   1.27 &   2.69 \\ 
   & (2)  & $6.81\times10^{-18}$ &   30.2 &   8.26 &  0.559 &  0.563 &   2.24 &   3.62 \\[4pt] 
 \multirow{2}{*}{WNM}  & (1)  & $4.17\times10^{-18}$ &   28.7 &   5.85 &  0.236 &  0.611 &   1.17 &   2.51 \\ 
   & (2)  & $4.44\times10^{-18}$ &   26.1 &   8.32 &  0.413 &  0.545 &   2.11 &   3.37 \\ 
\enddata
\tablecomments{(1): Disk-like grains. (2): Spherical grains.} 
\end{deluxetable}

\begin{figure}[htbp]
\begin{center}
\plotone{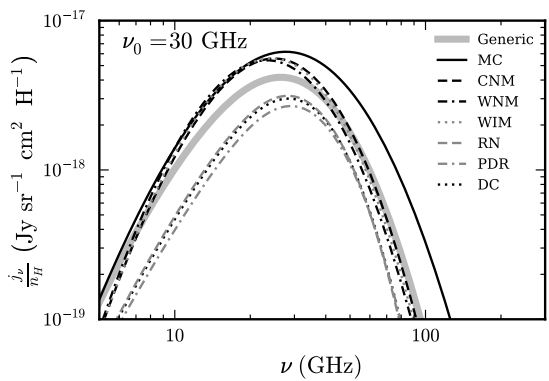}
\caption{$j_\nu/n_H$ curves for various environments, but with $\nu_0=30\,\mathrm{GHz}$, demonstrating the similar curve shapes.  The grey, solid curve shows the function using the suggested generic parameters.}
\label{j_nu_vs_nu_compare.fig}
\end{center}
\end{figure}

The parameters in Equation~\ref{j_nu_nh.eq} are not independent.  They depend on the excitation and damping power laws ($\alpha_a$, $\alpha_b$, and $\alpha_\nu$), the rotational peak $\Omega_{p,a_0}$, and the tumbling parameters ($R_0$, $q_r$, and $\sigma_r$), which are themselves dependent on the environment and grain properties.  Allowing the parameters of Equation~\ref{j_nu_nh.eq} to vary independently will complicate physical interpretation.  However, despite the ranges of these parameters, the $j_\nu/n_H$ curves show remarkably little diversity.  This can be seen by plotting the analytical curves for the various environments, but with $\nu_0$ set to some constant value.  This is done in Figure~\ref{j_nu_vs_nu_compare.fig}, with $\nu_0=30\,\mathrm{GHz}$.  It is apparent that the choice of generic values for $\alpha_s$, $\sigma_s$, $\eta_\nu$ and $\eta_a$ are likely to provide a satisfying fit in any environment.

The presence of the $a_m$ to $a_0$ ratio allows probing of the smallest grain size.  This requires care, though, as it will be heavily degenerate with $\nu_0$, itself depending on environment.  If the latter can be constrained independently, then measuring $a_m$ with this method will provide a new window on grain formation and destruction.

Judging from published data, breaking the degeneracies in this model will be challenging.  I suggest setting $a_m=a_0$, using the generic values for $\alpha_s$, $\sigma_s$, $\eta_\nu$ and $\eta_a$, and allowing only $\left.j_\nu/n_H\right|_{\nu_0}$ and $\nu_0$ to vary.  This is comparable to the analysis of \citet{Bennett:2012p685} and \citet{Planck:2013p686}.  As data quality improves, varying $\sigma_s$ may allow an improved fit.  If the fit is still unsatisfactory, then fitting the $\alpha_a$, $\alpha_b$, and $\alpha_\nu$ directly may be best, as the higher level parameters are ultimately functions of these.  Caution should be exercised, though, as inaccuracies in the model due to approximations may become significant at this point. 

\begin{figure}[htbp]
\begin{center}
\plotone{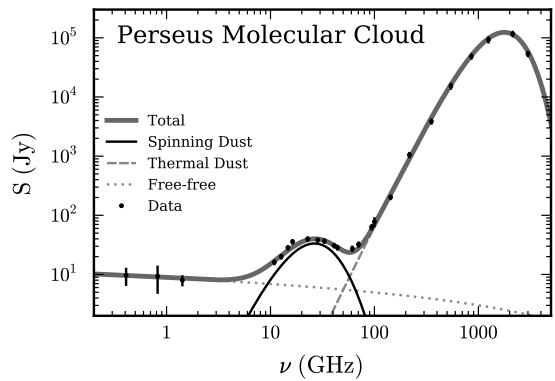}
\caption{Spectral energy distribution of the Perseus Molecular Cloud.  The data are as published in \cite{Planck:2011p338}, as are the free-free and thermal dust models.  The spinning dust model is from this work, with generic parameters, $\nu_0=30\,\mathrm{GHz}$, and scaled by a factor of $1.15$.}
\label{perseus_sed.fig}
\end{center}
\end{figure}

The model is demonstrated through comparison to the Perseus Molecular Cloud data of \citet{Planck:2011p338} in Figure~\ref{perseus_sed.fig}.  The free-free and thermal dust curves are taken directly from their fit.  The spinning dust model is that of this paper, with the generic parameters assumed and $\nu_0=30\,\mathrm{GHz}$.  The amplitude of the emission was scaled by $1.15$ to improve agreement.  The model provides a good fit to the data, with $\chi^2/\mathrm{d.o.f.}=0.81$.  This counts five parameters in the free-free, thermal dust, and CMB anisotropy \citep{Planck:2011p338} and the peak frequency and amplitude of the spinning dust model.

\section{Discussion}\label{discussion.sec}

The analytical derivation presented in this work allows one to understand spinning dust emission intuitively.  Emission from a given grain size is spread over a broad peak.  The breadth is greater if rotation is not limited by electric dipole damping, but is thermal.  The emission is also broadened if rotation is a function of the electric dipole moment, as in the case of plasma drag and electric dipole damping, though not enough to overcome non-thermal rotation.  There is a further broadening due to tumbling rotation.  Larger grains rotate more slowly, so integrating over grain size leads to a gently sloped, low frequency tail.  The existence of a smallest grain size leads to an exponential cut-off at high frequency.  The log-normal shape is largely due to the log-normal grain size distribution, though I have shown that emission from a given grain size is also well-approximated by this form.

As this derivation involved taking the products many independent functions, a log-normal shape is not surprising.  Indeed, simple algebra would allow factoring the power-law component of Equation~\ref{j_nu_nh.eq} into the log-normal component.  A pure log-normal spectrum is not justified, though, as the high frequency cut-off gives asymmetry to the spectrum.  This asymmetry is important theoretically, as it contains information on the grain size distribution, and observationally, because anomalous microwave emission measurements from \emph{Wilkinson Microwave Anisotropy Probe} and \emph{Planck} are at frequencies above the peak \citep{Bennett:2012p685,Planck:2013p686}.  

Different interstellar environments lead to different power laws in the distribution function and thus to different combinations of low-frequency slope, log-normal width, and high-frequency fall-off.  As seen in Figure~\ref{j_nu_vs_nu_compare.fig}, however, these effects do not lead to large deviations in spectral shape.  The characteristic frequency $\nu_0$ does vary with environment as it is closely related to the grain rotation temperatures.  There is degeneracy between $\nu_0$ and the other parameters of Equation~\ref{j_nu_nh.eq}, implying that the spectral shape ought to change as $\nu_0$ is shifted.  However, the similarity between curves in Figure~\ref{j_nu_vs_nu_compare.fig} suggests that such variations in shape will only become important when the precisions of anomalous microwave emission observations have greatly improved.

Measurements of the characteristic frequency $\nu_0$ will face strong degeneracy with the smallest grain size.  Decreasing $a_m$ gives smaller, faster rotating grains which extend the radiation to higher frequencies.  This will present a significant challenge to any attempts to constrain environment or smallest grain size with spinning dust radiation.

A number of approximations were needed in this work.  These are the power-law dependencies on $a$, $b$, and $\Omega$, the log-normal distribution function, and the log-normal spectrum for a tumbling grain.  Disagreement with the SpDust model below $\nu_0$ is mainly due to the first, while the latter two are to blame above $\nu_0$.  Transient effects due to individual gas collisions and the rotational consequences of triaxial grains and vibrational-rotational coupling were disregarded.  These omissions cause inaccuracy above $\nu_0$ and possible frequency shifts and broadening of the spectrum, respectively.

The caveats of this work extend beyond the analytical approximations.  Whether fitting numerical or analytical models, one must bear these in mind.  The most important are briefly discussed.
\begin{description}
\item[Grain size distribution.]  The log-normal form of this distribution is inspired by convenience, not astrophysics.  It is a four parameter model, with parameter values consistent with but not required by infrared and extinction data \citetalias{Weingartner:2001p526}.  These parameters may vary with local conditions, as may the form of the distribution itself.  Such variations would be degenerate with variations in rotational excitation and damping.
\item[Smallest grain size.]  Sublimation of small grains is a runaway process \citep{Guhathakurta:1989p536}.  Below a given size, grains will have a very short lifetime.  However, a sudden cutoff is not predicted: a range of grains sizes will be undergoing sublimation.  Further, sublimation is likely to dehydrogenate the grains before destroying the carbon skeleton, which will undoubtedly affect the electric dipole moments of these grains.   Spinning dust models ignore these complications, which may lead to structure above $\nu_0$.
\item[Grain geometry.]  A sharp transition between disk-like and spherical grains is unlikely to be physical.  Indeed, the very existence of purely disk-like and spherical grains is itself an approximation.  The true geometries of the grains will be more complex and could conspire to have grains of different sizes radiating at the same frequency.  This would cause structure in the spectrum below $\nu_0$.
\item[Electric dipole moments.]  \citetalias{Draine:1998p126} noted that the permanent electric dipole moments of these grains are extremely uncertain.  The random-walk prescription laid out therein and adopted thereafter leads to agreeable results, but is not physically motivated.  Possible effects can be probed by varying $\alpha_b$ and $\beta$.  An increase in the former broadens the spectrum via $\sigma_s$, but weakens it via $A_s$.  An increase in the latter brightens the emission, but boosts electric dipole damping to decrease $\nu_0$.  A breakdown in the random-walk prescription would introduce $a$ dependencies in $\alpha_b$ and $\beta$, potentially leading to structure in the spectrum.
\item[Rotational distribution function.]  The Fokker-Planck equation assumes that the damping and excitation mechanisms are stationary processes and that the impulses are small compared to the overall motion.  \citetalias{Hoang:2010p316} showed, however, that impulsive torques lead to a non-thermal tail at high angular momenta.  This translates to additional radiation above $\nu_0$ which the Fokker-Planck approach cannot capture.
\end{description}
These caveats may ultimately limit the precision of spinning dust models, as they can only be resolved through detailed knowledge of grain chemistry.  On the other hand, if future observations improve enough to require such precision, then constraining this chemistry directly may become possible.

I have described a new, analytical derivation of the radiation from spinning dust grains.  This work bypasses the lengthy, numerical calculations of previous models while encouraging an intuitive picture of the radiation.  Accuracy is not significantly compromised by this approach.  Indeed, strong caveats are present in even the numerical models.  These approximations are clearly described and their applicability is demonstrated by comparison with numerical calculations.  The final result is a simple, analytical function, well-suited for fitting to astronomical data.

\acknowledgements

I thank Y. Ali-Ha\:{i}moud, K. Cleary, C. Dickinson, B. Hensley, C. Hirata, T. Pearson, A. Readhead, C. Tibbs, and J. Villadsen for many useful conversations on the spinning dust radiation as this paper evolved.   I also thank the anonymous referee for many insightful suggestions.  This work was supported by the NSF grants AST-1010024 and AST-1212217.



\bibliographystyle{apj}
\bibliography{apj-jour,stevenson_0426}

\end{document}